\documentclass{JHEP3}
\usepackage{epsfig}
\usepackage{amsbsy}
\usepackage{varioref}
\usepackage{pifont}
\usepackage{axodraw}

\newcommand{\hs}{\hspace*{0.5cm}}

\newcommand{\be}{\begin{equation}}
\newcommand{\ee}{\end{equation}}
\newcommand{\bea}{\begin{eqnarray}}
\newcommand{\eea}{\end{eqnarray}}
\newcommand{\bary}{\begin{array}}
\newcommand{\eary}{\end{array}}
\newcommand{\bit}{\begin{itemize}}
\newcommand{\eit}{\end{itemize}}
\newcommand{\ben}{\begin{enumerate}}
\newcommand{\een}{\end{enumerate}}
\newcommand{\crn}{\nonumber \\}
\newcommand{\nn}{\nonumber}

\newcommand{\al}{\alpha}
\newcommand{\la}{\lambda}
\newcommand{\bet}{\beta}
\newcommand{\ga}{\gamma}

\newcommand{\fr}{\frac}

\newcommand{\bc}{\begin{center}}
\newcommand{\ec}{\end{center}}
\newcommand{\Ga}{\Gamma}
\newcommand{\de}{\delta}

\newcommand{\si}{\sigma}
\newcommand{\Si}{\Sigma}

\newcommand{\Om}{\Omega}

\def\sla#1{\ifmmode%
\setbox0=\hbox{$#1$}%
\setbox1=\hbox to\wd0{\hss$/$\hss}\else%
\setbox0=\hbox{#1}%
\setbox1=\hbox to\wd0{\hss/\hss}\fi%
#1\hskip-\wd0\box1 }
\title{Right-handed sneutrinos as self-interacting dark matter
in supersymmetric economical 3-3-1  model}
\author{ H. N. Long   \\ICRANet, P.le della Repubblica 10, 65100 Pescara, Italy\\  and
\\
  Institute of  Physics, VAST, P. O. Box
429, Bo Ho, Hanoi 10000, Vietnam\footnote{Permanent address}\\
 E-mail:
\email{hnlong@iop.vast.ac.vn}}

 \abstract{

In this work we show that the supersymmetric economical
$SU(3)_C\otimes SU(3)_L \otimes U(1)_X $(3-3-1) gauge model has a
realistic candidate for self-interacting dark matter. In the model
under consideration, the right-handed sneutrino is in bottom of
the triplet, which is a singlet of  the Standard Model $SU(2)_L$
group. In addition, the right-handed sneutrino is the lightest
slepton. By these properties, the right-handed sneutrino is weakly
interacting with the Standard Model and stable without
introduction  of  extra symmetry. From the Spergel-Steinhardt
condition, the typical mass limit $\leq 10 $ MeV  is derived. With
self-interacting coupling constant fixed by supersymmetry, this
limit is deduced without any approximation. The condition for
thermal generated self-interacting dark matter in the Universe is
also obtained. }


\begin{document}

\maketitle

\section{\label{intro}Introduction}

One of the themes of the history of physics has been the discovery
that the world familiar to us is only a tiny part of an enormous
and multi-faceted Universe. Over the past ten years, astronomers
have recognized that the stuff that we are made of accounts for
only 4\% of the total content of the Universe.

Until a few years ago, the more satisfactory cosmological
scenarios were ones composed of ordinary matter, cold dark matter
and a contribution associated with the  cosmological constant. To
be consistent with inflationary cosmology, the spectrum  of
density fluctuations would be nearly scale-invariant and
adiabatic. However, in recent years it has been pointed out that
the conventional models of collisionless cold dark matter (CCDM)
lead to problems with regard to galactic structures. N-body
simulations with CCDM indicate that galaxies should have singular
halos \cite{sinhalos} with large numbers of subhalos. The CCDM
predictions for the Tullly-Fisher relation and the stability of
galactic bars in high surface brightness spiral galaxies are not
in agreement with what is observed, indicating lower density
galaxy cores than predicted by CCDM. A number of other
inconsistencies, which we will not describe here, are discussed in
\cite{hon}.

In order to overcome the possible difficulties of CCDM, one
suggestion has been that the cold dark matter particles have a
non-dissipative self-interaction \cite{ss,ssmoi}, and it has been
shown that such cold, non-dissipative self-interacting dark matter
(SIDM) ~\cite{simdgood,SIDM} can be effective in alleviating the
various problems of CCDM \cite{tot}. One should notice that
self-interacting models lead to spherical halo centers in
clusters, which is not in agreement with ellipsoidal centers
indicated by strong gravitational lensing observations and by
Chanda ones. However, SIDM models are self-motivated as
alternative models. The key property of this kind of matter is
that, although its annihilation cross-section is suppressed, its
scattering cross section is enhanced.

Several authors have proposed models in which a specific scalar
singlet that satisfies the SIDM properties is introduced in the
Standard Model (SM) in an {\it ad hoc} way \cite{simdgood,SIDM}.
To be stable, this scalar cannot interact strongly with the SM
particles and it is guaranteed by introduction of an extra
symmetry (usually an $ U(1)$).

The first gauge model for SIDM were found by Fregolente and
Tonasse \cite{frog}  in the minimal 3-3-1 model. The next version
of SIDM is the 3-3-1 model with right-handed neutrinos
\cite{longlan} (For alternative direction in which the singlet
Higgs fields are WIMP, see Ref.~\cite{pires}).

 One of the main motivations to study the 3-3-1  models
is an explanation in part of the generation number puzzle. In the
3-3-1 models, each generation is not anomaly free; and the model
becomes anomaly free if one of quark families behaves differently
from other two \cite{ppf,flt}. Consequently, the number of
generations is multiple of the color number. Combining with the
QCD asymptotic freedom, the generation number has to be three.

In one of the 3-3-1 models, the right-handed neutrinos are in
bottom of the lepton triplets \cite{331rh} and three Higgs
triplets are required. It is worth noting that in the version with
right-handed neutrinos, there are two Higgs triplets with {\it
neutral components in the top and bottom}. In the earlier version,
these triplets can have vacuum expectation value (VEV) either on
the top or in the bottom, but not in both. Assuming that all
neutral components in the triplet can have VEVs, we are able to
reduce number of triplets in the model to be two
~\cite{ponce,haihiggs}. Such a scalar sector is minimal, therefore
it has been called the economical 3-3-1 model~\cite{higgseconom}.
In a series of papers, we have developed and proved that this
non-supersymmetric version is consistent, realistic and very rich
in physics \cite{haihiggs,higgseconom, dlhh,dls1}.

It is known that the economical (non-supersymmetric) 3-3-1 model
does not furnish any candidate ~\cite{higgseconom} for SIDM  with
the condition given by Spergel and Steinhardt~\cite{ss}. In the
other hands, supersymmety \cite{susy} contains interesting Higgs
physics \cite{shiggs}, where Higgs masses are constrained by
supersymmetry.  While earlier one might have viewed the Higgs
fields as just one of many features of low energy supersymmetric
models, the constraints on the Higgs mass are now problematic.
With a larger content of the scalar sector, the supersymmetric
version is expected to have a candidate for the self-interaction
dark matter. The scalar Higgs sector in the supersymmetric
economical 3-3-1 model does not provide the candidate for SIDM
\cite{susyeco}. In this paper, we show that the right-handed
sneutrinos are good candidates for the SIDM.

This paper is organized as follows. In Sec. \ref{model} we
recapitulate the necessary elements of   the  model under
consideration. The couplings of SIDM are presented in Sec.
\ref{int}, while in Sec. \ref{snlimit} we derive the lower  mass
limit for the SIDM. In Sec. \ref{cosmicdensity} we get the
condition for thermal generation of SIDM. Finally, the last
section - Sec. \ref{conc} is devoted to our conclusions.

\section{Basic elements}
\label{model} In this section we first recapitulate the basic
elements  of the model \cite{susyeco}, which are related to our
analysis below.

\subsection{\label{parcontent}Particle content }

 The superfield content in this paper is defined in a standard way as
follows \be \widehat{F}= (\widetilde{F}, F),\hs \widehat{S} = (S,
\widetilde{S}),\hs \widehat{V}= (\lambda,V), \ee where the
components $F$, $S$ and $V$ stand for the fermion, scalar and
vector fields while their superpartners are denoted as
$\widetilde{F}$, $\widetilde{S}$ and $\lambda$, respectively
\cite{susy,s331r}.

The superfields for the leptons under the 3-3-1 gauge group
transform as
\begin{equation}
\widehat{L}_{a L}=\left(\widehat{\nu}_{a}, \widehat{l}_{a},
\widehat{\nu}^c_{a}\right)^T_{L} \sim (1,3,-1/3),\hs
  \widehat {l}^{c}_{a L} \sim (1,1,1),\label{l2}
\end{equation} where $\widehat{\nu}^c_L=(\widehat{\nu}_R)^c$ and $a=1,2,3$
is a generation index. Here and in the following, the values in
the parentheses denote quantum numbers based on the
$\left(\mbox{SU}(3)_C,\mbox{SU}(3)_L,\mbox{U}(1)_X\right)$
symmetry.

The superfields for the left-handed quarks of the first generation
are in triplets \be \widehat Q_{1L}= \left(\widehat { u}_1,\
                        \widehat {d}_1,\
                        \widehat {u}^\prime
 \right)^T_L \sim (3,3,1/3),\label{quarks3}\ee
where the right-handed singlet counterparts are given by\be
\widehat {u}^{c}_{1L},\ \widehat { u}^{ \prime c}_{L} \sim
(3^*,1,-2/3),\hs \widehat {d}^{c}_{1L} \sim (3^*,1,1/3 ).
\label{l5} \ee Conversely, the superfields for the last two
generations transform as antitriplets
\[
\begin{array}{ccc}
 \widehat{Q}_{\alpha L} = \left(\widehat{d}_{\alpha},
 - \widehat{u}_{\alpha},
 \widehat{d^\prime}_{\alpha}\right)^T_{L}
 \sim (3,3^*,0), \hs \al=2,3, \label{l3}
\end{array}
\]
where the right-handed counterparts are in singlets
\begin{equation}
\widehat{u}^{c}_{\alpha L} \sim \left(3^*,1,-2/3 \right),\hs
\widehat{d}^{c}_{\alpha L},\ \widehat{d}^{\prime c}_{\alpha L}
\sim \left(3^*,1,1/3 \right). \label{l4}
\end{equation}
The primes superscript on usual quark types ($u'$ with the
electric charge $q_{u'}=2/3$ and $d'$ with $q_{d'}=-1/3$) indicate
that those quarks are exotic ones. The mentioned fermion content,
which belongs to that of the 3-3-1 model with right-handed
neutrinos \cite{331rh,haihiggs} is, of course,  free from anomaly.

The two superfields $\widehat{\chi}$ and $\widehat {\rho} $ are at
least introduced to span the scalar sector of the economical 3-3-1
model \cite{higgseconom}: \bea \widehat{\chi}&=& \left (
\widehat{\chi}^0_1, \widehat{\chi}^-, \widehat{\chi}^0_2
\right)^T\sim (1,3,-1/3), \crn \widehat{\rho}&=& \left
(\widehat{\rho}^+_1, \widehat{\rho}^0, \widehat{\rho}^+_2\right)^T
\sim  (1,3,2/3). \label{l8} \eea To cancel the chiral anomalies of
Higgsino sector, the two extra superfields $\widehat{\chi}^\prime$
and $\widehat {\rho}^\prime $ must be added as follows \bea
\widehat{\chi}^\prime&=& \left (\widehat{\chi}^{\prime 0}_1,
\widehat{\chi}^{\prime +},\widehat{\chi}^{\prime 0}_2
\right)^T\sim ( 1,3^*,1/3),\crn \widehat{\rho}^\prime &=& \left
(\widehat{\rho}^{\prime -}_1,
  \widehat{\rho}^{\prime 0},  \widehat{\rho}^{\prime -}_2
\right)^T\sim (1,3^*,-2/3). \label{l10} \eea In this model, the $
\mathrm{SU}(3)_L \otimes \mathrm{U}(1)_X$ gauge group is broken
via two steps:
 \be \mathrm{SU}(3)_L \otimes
\mathrm{U}(1)_X \stackrel{w,w'}{\longrightarrow}\ \mathrm{SU}(2)_L
\otimes \mathrm{U}(1)_Y\stackrel{v,v',u,u'}{\longrightarrow}
\mathrm{U}(1)_{Q},\label{stages}\ee where the VEVs are defined by
\bea
 \sqrt{2} \langle\chi\rangle^T &=& \left(u, 0, w\right), \hs \sqrt{2}
 \langle\chi^\prime\rangle^T = \left(u^\prime,  0,
 w^\prime\right),\crn
\sqrt{2}  \langle\rho\rangle^T &=& \left( 0, v, 0 \right), \hs
\sqrt{2} \langle\rho^\prime\rangle^T = \left( 0, v^\prime,  0
\right).\eea The VEVs $w$ and $w^\prime$ are responsible for the
first step of the symmetry breaking while $u,\ u^\prime$ and $v,\
v^\prime$ are for the second one. Therefore, they have to satisfy
the constraints:
 \be
 u,\ u^\prime,\ v,\ v^\prime
\ll w,\ w^\prime. \label{contraint}\ee

The vector superfields $\widehat{V}_c$, $\widehat{V}$ and
$\widehat{V}^\prime$ containing the usual gauge bosons are,
respectively, associated with the $\mathrm{SU}(3)_C$,
$\mathrm{SU}(3)_L$ and $\mathrm{U}(1)_X $ group factors. The
colour and flavour vector superfields have expansions in the
Gell-Mann matrix bases $T^a=\lambda^a/2$ $(a=1,2,...,8)$ as
follows\bea \widehat{V}_c &=& \fr{1}{2}\lambda^a
\widehat{V}_{ca},\hs
\widehat{\overline{V}}_c=-\fr{1}{2}\lambda^{a*}
\widehat{V}_{ca};\hs \widehat{V} = \fr{1}{2}\lambda^a
\widehat{V}_{a},\hs \widehat{\overline{V}}=-\fr{1}{2}\lambda^{a*}
\widehat{V}_{a},\nn\eea where an overbar $^-$ indicates complex
conjugation. For the vector superfield associated with
$\mathrm{U}(1)_X$, we normalize as follows \be X \hat{V}'= (XT^9)
\hat{B}, \hs T^9\equiv\fr{1}{\sqrt{6}}\mathrm{diag}(1,1,1).\ee In
the following, we are denoting the gluons by $g^a$ and their
respective gluino partners by $\lambda^a_{c}$, with $a=1,
\ldots,8$. In the electroweak sector, $V^a$ and $B$ stand for the
$\mathrm{SU}(3)_{L}$ and $\mathrm{U}(1)_{X}$ gauge bosons with
their gaugino partners $\lambda^a_{V}$ and $\lambda_{B}$,
respectively

\subsection{ Higgs content}

One of the most important things in study Higgs sector is
recognition the SM Higgs boson. Since it is  electrically neutral,
we are interested in only neutral Higgs bosons. Expansion of Higgs
fields in the model under consideration, is \cite{higph}
 \bea \chi^T&=&\left(
           \begin{array}{ccc}
            \fr{u+S_1+iA_1}{\sqrt{2}}, & \chi^{-}, & \fr{w+S_2+iA_2}{\sqrt{2}} \\
           \end{array}
         \right), \hs  \rho^T = \left(
 \begin{array}{ccc}
  \rho_1^+, & \fr{v+S_5+iA_5}{\sqrt{2}}, & \rho_2^+ \crn
          \end{array}
      \right),\label{2}\\ {\chi^\prime}^T&=&\left(
           \begin{array}{ccc}
\fr{u^\prime+S_3+iA_3}{\sqrt{2}}, & \chi^{\prime
+}, & \fr{w^\prime+S_4+iA_4}{\sqrt{2}} \\
           \end{array}
         \right),\hs {\rho ^\prime}^T =\left(
                         \begin{array}{ccc}
                           \rho_1^{\prime -}, &
\fr{ v^\prime +S_6+iA_6}{\sqrt{2}}, & \rho_2^{\prime-} \\
\end{array}\right).\label{1}
\eea

The weak eigenstates  and  physical eigenstates are related
through the following  matrix
 \bea \left(
 \begin{array}{c}
 S_1 \\
 S_2 \\
 S_3 \\
 S_4\\
 S_5\\
 S_6 \\
\end{array}
\right) &=&\left(
 \begin{array}{cccccc}
c_\beta s_\theta & -s_\beta c_\theta
& -c_\beta c_\theta & -s_\alpha s_\beta s_\theta &  -c_\alpha s_\beta s_\theta & 0 \\
 c_\beta c_\theta & s_\beta s_\theta & c_\beta s_\theta & -s_\alpha s_\beta c_\theta
 & -c_\alpha s_\beta c_\theta & 0 \\
s_\beta s_\theta & -c_\beta c_\theta & s_\beta c_\theta  &
s_\alpha c_\beta s_\theta
& c_\alpha c_\beta s_\theta & 0 \\
 s_\beta c_\theta & c_\beta s_\theta & -s_\beta s_\theta & s_\alpha c_\beta c_\theta
 & c_\alpha c_\beta c_\theta & 0 \\
 0& 0 & 0 & -c_\alpha c_\gamma & s_\alpha c_\gamma & s_\gamma \\
0 & 0 & 0 & c_\alpha s_\gamma & -s_\alpha s_\gamma & c_\gamma\\
\end{array}
\right)\left(
\begin{array}{c}
S'_{1a} \\
\varphi_{S_{24}}\\
\phi_{S_{24}}\\
H \\
 \phi_{S_{a36}} \\
 S'_5 \\
\end{array}
 \right)
\label{quidoi}
 \eea
\be t_{\theta}
\equiv \fr{u}{w}=\fr{u'}{w'}, \ t_{2\al}\equiv
\fr{-2m^2_{36a}}{m^2_{66a}-m^2_{33a}}\propto \fr{v}{w}, \ t_\bet
\equiv \frac{w}{w^\prime}, \ \cot_\gamma \equiv\frac{v}{v^\prime}.
\label{ht2tan}\ee
 Pursuing interactions of the scalar Higgs bosons
with the SM gauge ones, it was recognized that the following $H$
is the SM Higgs boson \cite{higph}: \bea H & = & s_\al S'_3 +c_\al
S'_6,\crn
 m^2_H & = & \frac{1}{2}
\left[m^2_{33a}+m^2_{66a}-
\sqrt{\left(m^2_{33a}-m^2_{66a}\right)^2 +4m^4_{36a}}
\right],\label{klvhnho} \eea where \bea m^2_{33a}&=&\fr{18 g^2 +
g'^2}{54c^2_\theta}(w^2 + w'^2),\hs m^2_{66a}=\fr{9g^2 +
2g'^2}{27}(v^2 + v'^2),\label{peca1021}\\
m^2_{36a}&=&\fr{(9g^2+2g'^2)\sqrt{(v^2 + v'^2)(w^2 + w'^2)}}{54
c_\theta}\label{peca1022}\eea
 From (\ref{peca1021}) and
(\ref{peca1022}), we have \be t_{\al} \propto \fr{v}{w} \
\Rightarrow  \ t_{\al} \gg t_\theta. \label{peca1062}\ee Taking
into account $\alpha=\fr{e^2}{4\pi}=\fr{1}{128},\ s_W^2=0.2312$,
we have \[
 m_H \simeq
 91.4 \  \textrm{GeV}.
\]
This value is very closed to the lower limit of 89.8 GeV (95\% CL)
given in Ref. \cite{pdg} p. 32. It is interesting to note that
this mass is also closed to the $Z$ boson mass.

\subsection{\label{nlepm}Right-handed sneutrinos - SIDM candidates }

In Ref.~\cite{susyeco}, we have introduced all of the possible
soft terms to break supersymmetry. As a result, our effective
Lagrangian of supersymmetric breaking is the most general. The
different sources of supersymmetric breaking such as
Fayet-Iliopoulos ($D$-term), O'Raifeartaigh  ($F$-term),
gauge-mediated,... lead to the  Lagrangian given in Eq. (18) of
Ref. \cite{susyeco}.

 In the previous work \cite{sfer}, we have shown
that the right-handed sneutrinos  are the lightest sfermions. Let
us remind some definitions. In the base ($\tilde{\nu}_{a L},
\tilde{\nu}_{b R})$=($\tilde{\nu}_{1 L}$, $ \tilde{\nu}_{2
L}$,$\tilde{\nu}_{3 L}$, $ \tilde{\nu}_{1 R}$, $\tilde{\nu}_{2
R}$, $\tilde{\nu}_{3 R}$), the mass
 matrix is given by \cite{sfer}
 \bea
\left(%
\begin{array}{cccccc}
  A_{ab} &  E_{ab} \\
  E_{ab} &  G_{ab}
\end{array}%
\right),
 \label{sf29} \eea
where \bea  A_{ab} & = & \fr{g^2}{2} \de_{a b}\left( N_3
 + \fr{1}{\sqrt{3}}N_8 - \fr{2 t^2}{3}N_1 \right)
  + M^2_{ab} + \fr 1 4 \mu_{0a} \mu_{0b}\label{durh1}\\ &&
  +\fr{1}{18}v^2(\la_a \la_b+4
\la^\prime_{ca}\la^\prime_{cb})+ \fr{1}{18}\la_a \la_b  w ^2,\crn
 G_{ab}& =&  -g^2 \de_{a b}\left( \fr{1}{\sqrt{3}}N_8 +
 \fr{ t^2}{3}N_1 \right) + M^2_{ab} + \fr 1 4 \mu_{0a} \mu_{0b}
 \crn &&+\fr{1}{18}v^2(\la_a \la_b+4
\la^\prime_{ca}\la^\prime_{cb})+ \fr{1}{18}\la_a \la_b  u^2,\crn
E_{ab} & = & - \sqrt{2}\left(\varepsilon_{ab} v +   \fr{1}{6
}\mu_\rho \la^\prime_{ab}v'\right),\label{sf29}\eea and
\cite{sfer}
 \bea N_3 & =&  -\fr 1 4 \left(u^2 \fr{\cos
2\bet}{s_\bet^2} + v^2 \fr{\cos 2\ga}{c_\ga^2} \right),\crn
 N_8 & = & \fr{1}{4\sqrt{3}} \left[ v^2 \fr{\cos
2\ga}{c_\ga^2}- (u^2 - 2 w^2)\fr{\cos 2\bet}{s_\bet^2}
\right],\crn N_1 & =  & \fr 1 6 \left[(u^2 +  w^2)\fr{\cos
2\bet}{s_\bet^2} + 2 v^2 \fr{\cos 2\ga}{c_\ga^2} \right].
\label{sf185}\eea
 As usual, we assume that  there is substantial  mixing among
$(\tilde{\tau}_L, \tilde{\tau}_R)$ {\it only} \cite{martin}. Then
 eigenstates and eigenmasses in this case are given
 in
Table~\ref{eigenn}.
\begin{table}[h]
\caption{Masses and eigenstates of sneutrinos}
\begin{center}
\begin{tabular}{|c|c|c|c|c|c|c|}
\hline
 Eigenstate &$\tilde{\nu}_{1 L}$ & $\tilde{\nu}_{2 L}$ & $\tilde{\nu}_{3
 L}$&
$\tilde{\nu}_{1 R}$ & $\tilde{\nu}_{2 R}$& $\tilde{\nu}_{3 R}$
\\
\hline
 $(\textrm{Mass})^2$ &$A_{11}$ & $A_{22}$ & $A_{33}$
 &$G_{11}$ & $G_{22}$  & $G_{33}$  \\
 \hline
\end{tabular}
\label{eigenn}
\end{center}
\end{table}

The mass splittings for the sleptons are governed by sum-rules
\cite{sfer} \bea m^2_{\tilde{l}_{1 L}} - m^2_{\tilde{\nu}_{1 L}} &
= & m^2_{\tilde{l}_{2 L}} - m^2_{\tilde{\nu}_{2 L}} = -  g^2 T_3 =
\fr{g^2}{4} \left( v^2 \fr{\cos 2\ga}{c_\ga^2} + u^2 \fr{\cos
2\bet}{s_\bet^2} \right)\crn & = & m_W^2 \cos 2 \ga + \fr{g^2
u^2}{4}\fr{\cos 2\bet}{s_\bet^2},
\label{sf201}\\
m^2_{\tilde{\nu}_{1 L}} - m^2_{\tilde{\nu}_{1 R}} & = &
m^2_{\tilde{\nu}_{2 L}} - m^2_{\tilde{\nu}_{2 R}} = \fr{g^2}{2}
\left( T_3 +\sqrt{3} T_8\right) =  \fr{g^2}{ 4} (w^2 -u^2)
\fr{\cos 2\bet}{s_\bet^2}.  \label{sf202}
 \eea
In the limit $u \approx 0$, Eq.(\ref{sf201}) is consistent with
those in the Minimal Supersymmetric Standard Model. Assuming
further $\cos 2 \bet
>0$, we obtain: $m^2_{\tilde{\nu}_{l L}}
> m^2_{\tilde{\nu}_{l R}}$.
Since no experimental data on supersymmetric partners, we have a
right to assume that.

To finish this section, we note that the right-handed sneutrinos
are the lightest sfermions (in company with suggestion $\cos 2
\bet >0$). So they are stable.  In addition, since they are
singlet of the SM $SU(2)_L$ gauge group, they do not interact with
the ordinary particles of the SM. For some range of the
parameters, they posse the right abundance for CDM (see below).
Hence they are realistic candidate for DM. Concerning
$\tilde{\nu}_{ a L}^{c}$ stability, notice that they carry  lepton
number $L = -1$, so final state of their decay must be slepton and
scalar Higgs boson. However, this is forbidden due to the
smallness of their  masses. For the short, let us call the
right-handed sneutrinos as dark matter and denote $\tilde{\nu}_{ a
L}^{c}$ by S.

\section{Interaction of the DM candidate }
\label{int}
 It is well-known that to be candidate for DM, particles
do not interact with the SM fields except, with the Higgs boson.
In the model under consideration, the couplings arise in both $F$-
and $D$-term  contributions. The scalar potential of the model is
a result of summation over $F$ and $D$ terms \cite{martin}: \be V
= F^{\phi *}F_\phi + \fr 1 2 \sum_a D^a D_a.
  \label{sf1}\ee
\ben \item  {\it Coupling from F-terms }

 Here we display only the $F$-terms giving necessary
interactions \cite{sfer}:
 \bea {\mathcal L}_F &=& \fr1 9 \la_a \la_b
[(\tilde{L}^*_{aL}\tilde{L}_{bL})(\rho^*\rho)
-(\tilde{L}^*_{aL}\rho)(\rho^*\tilde{L}_{bL})] \crn &&+ \fr1 9
\la_a \la_b [(\tilde{L}^*_{aL}\tilde{L}_{bL})(\chi^*\chi)
-(\tilde{L}^*_{aL}\chi)(\chi^*\tilde{L}_{bL})] \crn&&+\fr 4 9
\la^\prime_{ca}\la^\prime_{cb}[(\tilde{L}^*_{aL}\tilde{L}_{bL})(\rho^*\rho)
-(\tilde{L}^*_{aL}\rho)(\rho^*\tilde{L}_{bL})]\crn &&+ \fr 1 9
\ga_{ac} \ga_{b c} (\tilde{L}_{aL} \rho^{\prime}) (\tilde{L}_{b L}
\rho^{\prime})^*. \label{peca281}\eea Notations in this section is
given in Ref. \cite{sfer}.

 From (\ref{peca281}), we get couplings of
the right-handed sneutrinos with neutral scalar Higgs bosons: \bea
 {\mathcal L}^F_{SSHH}  &=&
 \fr1 9 \la_a \la_b (\tilde{\nu}_{ a
L}^{c *}\tilde{\nu}_{ b L}^{c})\left(\chi_1^{0*} \chi_1^{0} +
\chi_3^{0*} \chi_3^{0} + \rho^{0*} \rho^{0} \right) +\fr 4 9
\la^\prime_{ca}\la^\prime_{cb}(\tilde{\nu}_{ a L}^{c
*}\tilde{\nu}_{ b L}^{c})(\rho^{0*} \rho^{0} ).
 \label{peca282}\eea

It is worth noting that $\la_a$ is a coefficient of $R$-parity
violating interactions (see \cite{sfer}), hence they have to be
very small. Therefore, the main contribution in  (\ref{peca282})
is the last term. It was known that the mentioned term provides
mass for neutrinos, so it has to be much smaller as compared to
$\ga_{ac}$ \cite{dls1}: $\la^\prime_{ca} \ll \ga_{ac}$.

\item  {\it Coupling from D-terms }

As before, we display the terms giving necessary contribution
only. It also  exists in $D$-term forms: \bea D^a = -g
\left(\sum_{sfermions} \tilde{f}^\dag T^a \tilde{f} + \sum_{Higgs}
H^\dag T^a H \right).\label{sf181} \eea Since $T_a = T_a^\dag$, we
have \bea (D^a)^* D_a &= & \left(\sum_{sfermions} \tilde{f}^\dag
T^a \tilde{f}\right)^2 \crn  && + 2 g^2\left(\sum_{sfermions}
\tilde{f}^\dag T^a \tilde{f}\right) \left(\sum_{Higgs} H^\dag T^a
H \right)  + \cdot \cdot \cdot, \label{sf182} \eea where $\cdot
\cdot \cdot$ are the terms which do not contribute to sfermion
masses. The first term gives sfermion self-interactions. The
factor 2 in the second term in (\ref{sf182}) is the Newton's
binomial coefficient. Since sneutrino masses and interactions are
our interest, therefore, in the second factor at the last line of
(\ref{sf182}), only the diagonal $T_8$ satisfies this purpose.
This
 factor is given by: \bea H_8 & \equiv & \sum_{H=
\chi, \chi^\prime, \rho, \rho^\prime } <H^\dag > T_8 < H
>\crn
 &=& \fr{1}{2\sqrt{3}} \left(\chi_1^{0*} \chi_1^{0} - 2 \chi_3^{0*} \chi_3^{0} -
\chi_1^{' 0  *} \chi_1^{' 0} + 2 \chi_3^{' 0 *} \chi_3^{' 0} +
\rho^{0*} \rho^{0} - \rho^{' 0*} \rho^{' 0}
\right).\label{sf185}\eea Here we have taken into account that for
antitriplets, $T_8$ changes a sign. Let us  consider the first
factor of the about mentioned term in (\ref{sf182}). Since the
singlet fields do not give contribution, hence  for sleptons we
have: \bea
 SL_8  & \equiv  & \tilde{L}_{ a L}^\dag
T_8 \tilde{L}_{ a L}  =  \fr{1}{\sqrt{3}}\left(\fr 1 2
\tilde{\nu}_{ a L}^* \tilde{\nu}_{ a L} + \fr 1 2 \tilde{l}^*_{ a
L}\tilde{l}_{ a L} - \tilde{\nu}_{ a L}^{c *}\tilde{\nu}_{ a L}^{c
}\right).\label{sf187}\eea Thus, the contribution from $SU(3)_L$
subgroup is: \be g^2  SL_8 \times H_8. \label{sf192}\ee So,
sneutrino self-interaction arisen  from $ SL_8 $ is given by: \be
\fr{g^2}{6} (\tilde{\nu}_{ a L}^{c *}\tilde{\nu}_{ a L}^{c})^2.
\label{sf1041}\ee Now we are looking at $U(1)_X$ subgroup:

First, for the Higgs part, we have \bea H_1 & \equiv  &\sum_{H=
\chi, \chi^\prime, \rho, \rho^\prime } <H^\dag > X < H
> \crn & = & - \fr 1 3
[(\chi_1^{0*} \chi_1^{0} + \chi_3^{0*} \chi_3^{0}) - (\chi_1^{' 0
*} \chi_1^{' 0}  + \chi_3^{' 0  *} \chi_3^{' 0} ) - 2(\rho^{0*}
\rho^{0} -\rho^{' 0  *} \rho^{' 0} )].
 \label{sf194}\eea Similarly, for sleptons
\bea SL_1 & \equiv  & - \fr 1 3 (\tilde{\nu}^*_{ a L}
\tilde{\nu}_{ a L} + \tilde{l}^*_{ a L}\tilde{l}_{ a L} +
\tilde{\nu}^{c*}_{ a L} \tilde{\nu}^c_{ a L} ) + \tilde{l}^{c *}_{
a L}\tilde{l}^c_{ a L}.\label{sf195}\eea The contribution from
subgroup  $U(1)_X$  is \be
  g^{\prime 2}\times SL_1 \times H_1 =  g^2 t^2 \times SL_1 \times
  H_1.
 \label{sf197}\ee
Again, sneutrino self-interaction is given by \be \fr{g^2 t^2}{18}
(\tilde{\nu}_{ a L}^{c *}\tilde{\nu}_{ a L}^{c})^2,
\label{sf1042}\ee with  \cite{dl}\be
t=\fr{g'}{g}=\fr{3\sqrt{2}s_W}{\sqrt{4c^2_W-1}}.
 \ee
  The total contribution is a result of summation over
two above mentioned subgroup parts. Thus, the dark matter - Higgs
boson interactions  are given by
 \bea {\mathcal L}^D_{SSHH} & \in &  (SL_8 . H_8 + SL_1 . H_1)\crn
 &=& - \fr{g^2}{6}(\tilde{\nu}_{ a L}^{c *}\tilde{\nu}_{ a
L}^{c})(\chi_1^{0*} \chi_1^{0} - 2 \chi_3^{0*} \chi_3^{0} -
\chi_1^{' 0  *} \chi_1^{' 0} + 2 \chi_3^{' 0 *} \chi_3^{' 0} +
\rho^{0*} \rho^{0} - \rho^{' 0*} \rho^{' 0})\crn & & +
 \fr{g^2 t^2}{9} (\tilde{\nu}_{ a L}^{c *}\tilde{\nu}_{ a
L}^{c})[(\chi_1^{0*} \chi_1^{0} + \chi_3^{0*} \chi_3^{0}) -
(\chi_1^{' 0  *} \chi_1^{' 0}  + \chi_3^{' 0  *} \chi_3^{' 0} ) -
2(\rho^{0*} \rho^{0} -\rho^{' 0  *} \rho^{' 0} )].\crn
 \label{peca3} \eea
\een Hence the total DM-Higgs interaction Lagrangian is the
following \bea {\mathcal L}_{int} & = &  {\mathcal L}^F_{SSHH} +
{\mathcal L}^D_{SSHH}\crn
 & = & \fr1 9 \la_a \la_b (\tilde{\nu}_{ a
L}^{c *}\tilde{\nu}_{ b L}^{c})\left(\chi_1^{0*} \chi_1^{0} +
\chi_3^{0*} \chi_3^{0} + \rho^{0*} \rho^{0} \right) +\fr 4 9
\la^\prime_{ca}\la^\prime_{cb}(\tilde{\nu}_{ a L}^{c
*}\tilde{\nu}_{ b L}^{c})(\rho^{0*} \rho^{0} )\crn &&  -
\fr{g^2}{6}(\tilde{\nu}_{ a L}^{c *}\tilde{\nu}_{ a
L}^{c})(\chi_1^{0*} \chi_1^{0} - 2 \chi_3^{0*} \chi_3^{0} -
\chi_1^{' 0  *} \chi_1^{' 0} + 2 \chi_3^{' 0 *} \chi_3^{' 0} +
\rho^{0*} \rho^{0} - \rho^{' 0*} \rho^{' 0})\crn & & +
 \fr{g^2 t^2}{9} (\tilde{\nu}_{ a L}^{c *}\tilde{\nu}_{ a
L}^{c})[(\chi_1^{0*} \chi_1^{0} + \chi_3^{0*} \chi_3^{0}) -
(\chi_1^{' 0  *} \chi_1^{' 0}  + \chi_3^{' 0  *} \chi_3^{' 0} ) -
2(\rho^{0*} \rho^{0} -\rho^{' 0  *} \rho^{' 0} )].\crn
 \label{peca1066} \eea

Substitution of (\ref{1}) into (\ref{peca1066}) yields quartic
couplings\bea {\mathcal L}_{SSHH} &=&
 \fr{1}{18} \la_a \la_b (\tilde{\nu}_{ a
L}^{c *}\tilde{\nu}_{ b L}^{c})\left(S_1^2 + S_2^2 + S_5^2 + A_1^2
+ A_2^2 + A_5^2 \right)\crn &&+\fr{ 4}{18}
\la^\prime_{ca}\la^\prime_{cb}(\tilde{\nu}_{ a L}^{c
*}\tilde{\nu}_{ b L}^{c}) (S_5^2 + A_5^2)\crn &&  -
\fr{g^2}{12}(\tilde{\nu}_{ a L}^{c *}\tilde{\nu}_{ a L}^{c})(S_1^2
- 2 S_2^2 - S_3^2 + 2 S_4^2 + S_5^2 - S_6^2 \crn && + A_1^2 - 2
A_2^2 - A_3^2 + 2 A_4^2 + A_5^2 - A_6^2)\crn
 & & +
 \fr{g^2 t^2}{18} (\tilde{\nu}_{ a L}^{c *}\tilde{\nu}_{ a
L}^{c})[(S^2_1 + S^2_2) - (S^2_3 + S^2_4 ) - 2(S^2_5 -S^2_6 )\crn
&& + (A^2_1 + A^2_2) - (A^2_3 + A^2_4 ) - 2(A^2_5 - A^2_6 )].
 \label{peca27} \eea
We remind that   $A_5$, $ A_6$ are Goldstone bosons (massless)
\cite{susyeco} and three  massless states are mixing of \bea
A_1^\prime & = & s_\bet A_1 -c_\bet  A_3, \crn
 A_2^\prime & = & s_\bet A_2 - c_\bet A_4,\crn
\varphi_A&=& s_\theta A_3^\prime +c_\theta A_4^\prime,
 \label{giavohuong1t} \eea
where \be
  A_3^\prime = c_\bet
A_1 + s_\bet A_3,\hs A_4^\prime = c_\bet A_2 + s_\bet  A_4.
\label{candDM}\ee
 One massive eigenstate
\be \phi_A= c_\theta A_3^\prime - s_\theta
A_4^\prime\label{candDM2},\ee with mass  equal to those of the $X$
bilepton \cite{susyeco} \bea
m^2_{\phi_A}=\frac{g^2}{4}(1+t^2_\theta)(w^2+w^{\prime 2})=
m^2_{X}. \label{hx}\eea

Expressing $S_i, A_i, i=1,2,...,5$ through physical fields by
 (\ref{quidoi}), we will get quartic DM-DM-Higgs-Higgs
 interactions. However, we are just interested in the coupling
 of the SM  Higgs boson $H$. It reads
 \bea
 {\mathcal L}_{SSHH}  &=&\fr{1}{18} \la_a \la_b (\tilde{\nu}_{ a
L}^{c *}\tilde{\nu}_{ b L}^{c})H^2(s^2_\al s^2_\bet + c^2_\al
c^2_\ga) + \fr{ 4}{18}
\la^\prime_{ca}\la^\prime_{cb}(\tilde{\nu}_{ a L}^{c
*}\tilde{\nu}_{ b L}^{c})H^2  (c^2_\al c^2_\ga)\crn && +
\fr{g^2}{12}(\tilde{\nu}_{ a L}^{c *}\tilde{\nu}_{ a L}^{c})H^2 [
s^2_\al (1 - 3 c^2_\theta)(1-2 c^2_\bet)  + c^2_\al (1-2
s^2_\ga)\crn && - \fr{2 t^2}{3}(s^2_\al c_{2 \bet} + 2c^2_\al c_{2
\ga} )]. \label{peca284}\eea

Expression in (\ref{peca284}) can be rewritten in the form \bea
 {\mathcal L}_{SSHH}  &=&\fr{\la_{S  ab}}{18} (\tilde{\nu}_{ a
L}^{c *}\tilde{\nu}_{ b L}^{c})H^2, \label{peca286}\eea where \bea
\la_{S  ab}  &=&  \la_a \la_b (s^2_\al s^2_\bet + c^2_\al c^2_\ga)
+ 4  \la^\prime_{ca}\la^\prime_{c b} c^2_\al c^2_\ga \crn && +
\fr{3 \de_{a b} g^2}{2} [ s^2_\al (1 - 3 c^2_\theta)(1-2 c^2_\bet)
+ c^2_\al (1-2 s^2_\ga)\crn && - \fr{2 t^2}{3}(s^2_\al c_{2 \bet}
+ 2c^2_\al c_{2 \ga} )]. \label{peca1071}\eea

We turn now to the triple DM-DM Higgs boson interaction.
Substitution (\ref{1}) into  (\ref{peca1066}) yields \bea
{\mathcal L}_{SSH} &=&
 \fr1 9 \la_a \la_b (\tilde{\nu}_{ a
L}^{c *}\tilde{\nu}_{ b L}^{c})\left(u S_1 + w S_2 + v S_5
\right)\crn &&+\fr 4 9
\la^\prime_{ca}\la^\prime_{cb}(\tilde{\nu}_{ a L}^{c
*}\tilde{\nu}_{ a L}^{c})(v S_5 )\crn &&
 - \fr{g^2}{6}(\tilde{\nu}_{ a L}^{c
*}\tilde{\nu}_{ a L}^{c})(u S_1 - 2 wS_2 -u'S_3 + 2 w'S_4 + v S_5
- v'S_6)\crn && +\fr{g^2t^2}{9}(\tilde{\nu}_{ a L}^{c
*}\tilde{\nu}_{ a L}^{c})(u S_1 +  wS_2 -u'S_3 -  w'S_4 - 2 v S_5
+ 2 v'S_6). \label{peca1072}
 \eea

Expressing $S_i, i= 1,2,3,...,6$ through physical Higgs fields by
(\ref{1}) yields the necessary couplings. Then, we can write
triple DM-DM-Higgs couplings in the form: \bea {\mathcal L}_{SSH}
&=& \la_H H(\tilde{\nu}_{ a L}^{c *}\tilde{\nu}_{ b L}^{c}),
\label{peca1076}
 \eea
 where
 \bea
\la_H &=& - \fr{1}{9} \left[  \la_a \la_b (u s_\al s_\bet s_\theta
+ w s_\al s_\bet c_\theta + v  c_\al c_\ga) + 4
\la^\prime_{ca}\la^\prime_{cb} v c_\al c_\ga \right] \crn && +
\fr{\de_{a b}  g^2 }{6} \left[ 2 w \fr{s_\al c_\theta}{s_\bet} - v
\fr{c_\al }{c_\ga} - \fr{t^2}{3}\left( w \fr{s_\al
c_\theta}{s_\bet} -2 v \fr{c_\al }{c_\ga}\right)\right].
 \label{peca1077}
 \eea
Here we have taken into account $u \simeq u'$ \cite{higph}. As
mentioned above, both kinds of couplings constants in the
$F$-terms are small. Thus, the main contribution in
(\ref{peca1077}) is one from the $D$-terms.

The $D$-terms give also  dark matter self-interaction.
 This kind of
interaction exists only in $D$-terms.  Summation over
(\ref{sf1041}) and (\ref{sf1042}) yields
 quartic DM  self-interaction  \bea
 {\mathcal L}_{SSSS}  &=&  \fr{g^2}{6}(\tilde{\nu}_{ a
L}^{c *}\tilde{\nu}_{ b L}^{c})(\tilde{\nu}_{ b L}^{c
*}\tilde{\nu}_{ a L}^{c})\left(1+ \fr{t^2}{3}\right).
\label{peca288}\eea

Next, we turn on application of the above mentioned interactions
to physical processes relevant to the SIDM.

\section{Limit on sneutrino mass}
\label{snlimit}

 With self-interaction in (\ref{peca288}) we can get a limit for DM
 mass.  The  Spergel-Steinhardt condition
  on self-interaction cross-section of
  $S S^+ \rightarrow   S S^+$  has a
  form \cite{ss,simdgood}
   \be r_S = \fr{\si}{M } = (2.05
\times 10^{3}\  \div  \ 2.57 \times 10^4) \ \textrm{GeV}^{-3}.
\label{sscond} \ee From (\ref{peca288}), it follows that
 \be \si(S
S^+ \rightarrow   S S^+) + \si(S S \rightarrow   S S)= \fr{3}{128
\pi m_S^2}\left[\fr{2g^2}{3}\left(1 +
 \fr{t^2}{3}\right)\right]^2.
\label{peca1052} \ee
 Combination of  (\ref{sscond}) and
(\ref{peca1052}) implies that \be m_S = 35.8
\al_\eta^{1/3}\left(\fr{2.05 \times 10^3 \ \textrm{GeV}^{-3}}{r_S}
\right)^{1/3} \ \textrm{MeV} \label{peca101}\ee where \be \al_\eta
= \fr{1}{9 \pi}g^4 \left(1+ \fr{t^2}{3}\right)^2 =  \fr{16 m_W^4}{
9 \pi v^4} \left(1+ \fr{t^2}{3}\right)^2 = 0.027. \ee Here we have
used  $ m_W = 80.388 \ \textrm{GeV}, \ v = 246$ GeV \cite{pdg}.
Note that $\al_\eta$ in the model under consideration is quite
fair for perturbative theory and this is in good agreement with
estimation in Ref. \cite{simdgood}. Thus
 \be m_S =
\al_\eta^{1/3}(15.4 - 35.8) \ \textrm{MeV} \ \simeq (9\  \div \ 22
) \ \textrm{MeV}. \label{peca1011} \ee
 So sneutrino mass limit is
in the Spergel-Steihardt mass range $\sim 30$ MeV \cite{simdgood}.

\section{Thermal generation of self-interacting dark matter }
\label{cosmicdensity}

 The cosmic density of light gauge singlet scalars has been
 calculated in Ref. \cite{simdgood} and is given by
 \be
 \Om_H = 2 g (T_\ga) T^3_\ga \fr{\Si_i m_i \Theta_i}{\rho_c g(T)}
\label{peca1054} \ee with \be \Theta_i \equiv \fr{n_i}{T^3} =
\fr{\eta \Ga_i^2}{4 \pi^3 K m_H^3} \label{peca1055} \ee where
$T_\ga = 2.4 \times 10^{-4}$ eV is the present photon temperature,
$g (T_\ga) = 2$ is the photon degree of freedom, $g(T)= g_B + \fr
7 8 g_F$ ($g_B$ and $g_F$ are the relativistic boson and fermion
degree of freedom, respectively), $\rho_c = 7.5 \times 10^{-47}\
h^2 \ \textrm{GeV}^4$ is the critical density of the Universe ($h
\simeq 0.71$ is the Hubble constant in units of 100 km $s^{-1}
Mpc^{-1}$), $\eta = 1.87$, $K^2 = 4 \pi^3 g(T)/ 45 m^2_{pl}$ and
$m_{pl} = 1.2 \times 10^{19}$ GeV is the Planck mass. For
non-supersymmetric  3-3-1 model $g(T) \simeq 130$ ~\cite{frog},
and for the supersymmetric one, following  Ref.~\cite{macd2}, we
take $g(T) \simeq 260$. We will take $T= m_S$, since most of the
contribution to each $\Theta_i$ comes from $T \leq m_H \leq
T_{ew}$ \cite{simdgood}, where $T_{ew } \geq 1.5\ m_H$
\cite{jansen}.

Decay rate for $H \rightarrow S S^+$   is \be \Ga_H =
\fr{\la_H^2}{16 \pi m_H},\  \textrm{for}\  m_H \gg m_S,
\label{peca1095}\ee where $m_H, m_S $ are masses of  Higgs boson
and DM, respectively.

Numerical estimation yields \bea \la_H \approx 1.89 \times 10^{-6}
\left( \fr{g(T)}{260}\right)^{\fr 3 8} \left( \fr{m_H}{90\
\textrm{GeV}}\right)^{\fr 5 4} \left( \fr{m_S}{30\
\textrm{MeV}}\right)^{-\fr 1 4}.\label{durh2}\eea  Note that at
the tree level, the SM  Higgs boson has mass
 \be  m^2_H
\simeq (0.206 - 0.067/c^2_\theta) (v^2+v'^2) = (0.206 -
0.067/c^2_\theta)\  v^2_{SM}, \label{kln3} \ee where $ v_{SM} =
246$ GeV. Taking into account the upper limit~\cite{haihiggs}:
$\sin^2 \theta \leq 0.0064$, we get a mass of  the SM Higgs boson
at the tree level: $m_H \approx 91.573$ GeV. It is expected that
the radiative correction will give positive contribution to the
Higgs boson mass.

For the right-handed sneutrinos, we have \be m_S = (A_{11})^{\fr 1
2},\label{dur27}\ee where $A_{aa}$ is given by (\ref{durh1}).

Combining (\ref{peca1077})  and  (\ref{durh2}) yields \bea && -
\fr{1}{9} \left[  \la_a \la_b (u s_\al s_\bet s_\theta + w s_\al
s_\bet c_\theta + v  c_\al c_\ga) + 4
\la^\prime_{ca}\la^\prime_{cb} v c_\al c_\ga \right] \crn && +
\fr{2 \de_{a b}  m_W^2 }{3 v^2_{SM}} \left[ 2 w \fr{s_\al
c_\theta}{s_\bet} - v \fr{c_\al }{c_\ga} - \fr{t^2}{3}\left( w
\fr{s_\al c_\theta}{s_\bet} -2 v \fr{c_\al
}{c_\ga}\right)\right]\crn && =
 1.89 \times
10^{-6} \left( \fr{g(T)}{260}\right)^{\fr 3 8} \left( \fr{m_H}{90\
\textrm{GeV}}\right)^{\fr 5 4} \left( \fr{m_S}{30\
\textrm{MeV}}\right)^{-\fr 1 4}.\label{durh3}\eea

By suitable choice, the  condition (\ref{durh3}) for the SIDM  in
the model under consideration can be easily satisfied. Thus, a
system of three equations (\ref{kln3}), (\ref{dur27}) and
(\ref{durh3}) are the constraint conditions to guarantee that the
SIDM does not overpopulate the Universe.

\section{Conclusion} \label{conc} In this paper we have shown
that  the supersymmetric economical 3-3-1 model has  natural
candidates for the SIDM. It is the light right-handed sneutrinos.
The reason behind this choice relies on the fact that the
right-handed sneutrinos are  singlets of the SM $SU(2)_L$ group
and the lightest slepton. The first reason prevents interactions
of the DM candidates with particles in the SM, except for the
Higgs boson $H$. The second one stabilizes  the DM without
imposition extra symmetry.

In difference with the previous SIDM candidates which are scalar
Higgs bosons , the right-handed sneutrinos in this case are
superpartners of leptons with  $L = -1$. It is interesting to note
that in Ref.~\cite{xue}, the right-handed neutrinos are a possible
candidate  of {\it warm} dark matter.

  In order to be able to account for the
  observed properties of dark matter halos
(the Spergel-Steinhardt condition), the right-handed sneutrinos
have to be light with mass of ten MeV. It is emphasized that the
DM self-interaction is fixed from $D$-terms, hence  the above
mentioned limit was obtained without any assumption.
 Meanwhile they do not overpopulate the Universe with $\Om_H
= 0.3$. This dark matter arises naturally in the model without
imposition of extra symmetry.

Finally, we would like to mention that the economical 3-3-1 model
contains the minimal Higgs sector (economical) with very rich
phenomenology, specially in neutrino sector. Its supersymmetric
generalization has almost the same properties such as Higgs sector
and is very constrained. In addition, in this supersymmetric
version, the candidates for self-interacting dark matter exist
naturally.

\section*{Acknowledgments}
The author would like   thank  J. McDonald for useful comments. He
would like to express sincere gratitude to Prof. R. Ruffini for
invitation, S. S. Xue for discussion,   ICRANet for financial
support and its Members for hospitality during his visit.  This
work was supported in part by National Council for Natural
Sciences of Vietnam under grant No: 410604.
\\[0.3cm]

\end{document}